\documentclass[aps,prl,twocolumn,showpacs]{revtex4-1}

\usepackage{amssymb}
\usepackage{graphicx}
\usepackage{amsmath}
\usepackage{amsbsy}
\usepackage{amsthm}
\usepackage{bbm}
\usepackage{bm}
\usepackage{epsfig}
\usepackage{epstopdf}
\usepackage{dsfont}
\usepackage{hyperref}
\usepackage{soul}
\usepackage{color}

\newcommand{\be}{\begin{equation}}
\newcommand{\ee}{\end{equation}}
\newcommand{\beq}{\begin{eqnarray}}
\newcommand{\eeq}{\end{eqnarray}}

\DeclareMathOperator{\atanh}{atanh}

\DeclareMathOperator{\diag}{diag}

\newcommand{\BbbZ}{\mathbb{Z}}

\usepackage{geometry}
\begin{document}

\title{Motion generates entanglement}
\date{April 2012}

\author{Nicolai Friis}
\email{pmxnf@nottingham.ac.uk}
\author{David Edward Bruschi}
\email{pmxdeb@nottingham.ac.uk}
\author{Jorma Louko}
\email{jorma.louko@nottingham.ac.uk}
\author{Ivette Fuentes}
\email{ivette.fuentes@nottingham.ac.uk}
\thanks{previously known as Fuentes-Guridi and Fuentes-Schuller.}
\affiliation{School of Mathematical Sciences,
University of Nottingham,
University Park,
Nottingham NG7 2RD,
United Kingdom}
\begin{abstract}
We demonstrate entanglement generation between mode pairs of a quantum
field in a single, rigid cavity that moves nonuniformly in Minkowski
space-time. The effect is sensitive to the initial state, the choice
of the mode pair and bosonic versus fermionic statistics, and it can
be stronger by orders of magnitude than the entanglement degradation
between an inertial cavity and a nonuniformly moving cavity.
Detailed results are given for massless scalar and spinor fields in
$(1+1)$ dimensions. By the equivalence principle, the results model
entanglement generation by gravitational effects.
\end{abstract}

\pacs{
03.67.Mn,
03.65.Yz,
04.62.+v}

\maketitle

In the emerging field of relativistic quantum information,
the past decade has revealed a number of novel kinematic
effects on entanglement
between observers in relative inertial motion
\cite{inertial} and between one inertial and
one uniformly linearly
accelerating observer
(see Refs.~\cite{alsingmilburn03,fuentesschullermann05,alsingfuentesschullermanntessier06,
bruschiloukomartinmartinezdraganfuentes10,martinmartinezfuentes11,friiskoehlermartinmartinezbertlmann11,
smithmann11}
for a selection and Ref.~\cite{MartinMartinez:2011mw} for a recent review).
The phenomenon that has received most attention is the
entanglement \emph{degradation\/} caused by relative acceleration,
assuming that the issues of switching the acceleration
on and off are negligible and field modes of plane wave type,
spread over large regions of the space-time, are
considered.
A~rare case of entanglement generation
under these assumptions
is found in Ref.~\cite{arXiv:1011.6540}.

A framework that removes the idealisation
of spatially unlocalised field modes is to
confine the field to a cavity \cite{DownesFuentesRalph2011},
and the idealisation of everlasting acceleration can be removed
by taking the cavity's worldtube to be
inertial outside a finite time interval \cite{bruschifuenteslouko11}.
This creates a situation that should be more directly experimentally accessible.
When the cavity's worldtube consists of inertial and
uniformly linearly accelerated segments,
it was shown in
Refs.~\cite{bruschifuenteslouko11,friisleebruschilouko11}
that the evolution of the quantum field
is amenable to a systematic perturbative analysis
when the accelerations are small compared
with the cavity mode frequencies,
both for bosonic
and fermionic fields;
further, the perturbative analysis remains valid for arbitrary
durations of the individual segments, and the
distances travelled may hence be arbitrarily large.
The nonuniformly accelerated motion
was found to degrade mode entanglement
between the moving cavity and an inertial
reference cavity. The physical mechanism
behind the degradation is that
when observations of the field are
restricted to a small number of field modes,
for example by appealing to the frequency
sensitivity profile of a detector by which the field might be observed,
particle creation within the moving cavity causes
information to be lost into the infinitely many field
modes that are not being observed.

In this article we show, within the above perturbative
treatment of a quantum field that is confined
to a single cavity,
that \emph{nonuniform motion of a rigid cavity
creates entanglement between any two given
field modes within the cavity\/},
even when the initial state is separable.
We develop a general quantitative analysis for both
a scalar field and a fermion field,
giving
detailed results for a~sample travel scenario and
demonstrating that the particle statistics has a significant
effect on the entanglement.
As a highlight, we show that the
\emph{entanglement generation can appear in the first order
in the small acceleration expansion\/},
while the entanglement degradation between cavities is only a
second-order effect~\cite{bruschifuenteslouko11,friisleebruschilouko11}.
This suggests that entanglement generation
in a cavity may be more readily observable than
the entanglement degradation between cavities,
allowing more effective tests of phenomena
linked to the dynamical Casimir effect~\cite{dodonovCasimirReview2010},
and possibly also improving the technological prospects of
building quantum gates based on acceleration effects.
For instance, highly entangled two-mode squeezed states,
produced by a known gate in continuous variable systems, can be generated
by periodically repeating segments of uniformly accelerated and inertial
motion \cite{bruschidraganleefuenteslouko12}.


We work in $(1+1)$-dimensional Minkowski
space with metric signature $({-}{+})$:
additional transverse dimensions can be included via
their contribution to the effective field mass.
The length of the cavity in
its instantaneous rest frame is $\delta>0$.
The cavity is assumed to be inertial
outside a finite time interval, but the initial and
final velocities need not coincide.
We use units in which $\hbar=c=1$.
Complex conjugation is denoted by an asterisk
and Hermitian conjugation by a dagger.
$O(x)$ denotes a quantity for which
$O(x)/x$ is bounded as $x\to0$.

\textbf{Bosons}:\
We consider a real scalar field $\phi$ satisfying the
Klein-Gordon equation
$(-\Box+m^{2})\phi=0$,
where $m\ge0$ is the mass and $\Box$ is the scalar D'Alambertian.
For definiteness, we adopt at the cavity
boundaries the Dirichlet boundary condition
as in~\cite{bruschifuenteslouko11}, although
much of the analysis holds for any boundary condition that ensures
unitarity of the time evolution.

Let $\left\{\phi_{n} \mid n = 1,2,\ldots\right\}$ be a complete
orthonormal set of mode solutions that are of positive frequency
with respect to the cavity's proper time at early times (the in-region),
and let
$\bigl\{\tilde\phi_{n} \mid n = 1,2,\ldots\bigr\}$
be a similar set at late times (the out-region).
Each set has an associated set of ladder operators,
with the respective commutation relations
$\left[a_{n},a_{m}^{\dagger}\right]=\delta_{nm}$
and $\left[\tilde{a}_{n},\tilde{a}_{m}^{\dagger}\right]=\delta_{nm}$,
and a vacuum state, denoted respectively by $\left|\,0\,\right\rangle$
and $\left|\,\tilde{0}\,\right\rangle$.
The two sets of modes are related by the
Bogoliubov transformation
\begin{align}
\tilde{\phi}_{m}    &=  \sum\limits_{n}\left(\alpha_{mn}\phi_{n}+\beta_{mn}\phi_{n}^{*}\right)\,,
\label{eq:Bogo transformations phi phi-tilde}
\end{align}
and the ladder operators are related by
\begin{align}
a_{n} =  \sum\limits_{m}\left(\alpha_{mn}\tilde{a}_{m}+\beta_{mn}^{\,*}\tilde{a}_{m}^{\dagger}\right),
\label{eq:Bogo transformation operators out to in region}
\end{align}
where the notation is as in Ref.~\cite{birrelldavies}.
The vacua are related by \cite{fabbrinavarrosalas}
\begin{align}
\left|\,0\,\right\rangle=N\,e^{W}\left|\,\tilde{0}\,\right\rangle\,,
\label{eq:bosonic vacuum relation}
\end{align}
where $W:=\tfrac{1}{2}\sum_{pq}V_{pq}\tilde{a}_{p}^{\dagger}\tilde{a}_{q}^{\dagger}$,
$V := - \beta^{\,*} \alpha^{-1}$
and $N$ is a normalisation constant.

We prepare the system in the in-region in a state without mode
entanglement. We ask: does the cavity's motion generate mode
entanglement in the out-region, where the particle content of the
state has changed?

Our methodology is as follows.  We first specify the in-region state
and express it in the out-region basis using
(\ref{eq:bosonic vacuum relation}) and the adjoint
of~(\ref{eq:Bogo transformation operators out to in region}).
We then trace over all out-region modes except
those labelled by two distinct quantum numbers $k$ and~$k'$.  We
quantify the entanglement in the resulting reduced density matrix by
the negativity~\cite{peres96,horodeckiMPR96,vidalwerner02}, defined as
minus the sum of the negative eigenvalues of the partial transpose.
The advantages of negativity are that it is easy to compute and it
interpolates between entanglement monotones that have a more direct
operational interpretation~\cite{plenio-virmani:review}.

To identify a parameter regime that is treatable analytically,
we assume that the Bogoliubov coefficients
have Maclaurin expansions in a small dimensionless parameter~$h$, such
that
\begin{subequations}
\label{eq:alphas and betas small h expansion}
\begin{align}
\alpha  &=  \alpha^{(0)} + \alpha^{(1)} + \alpha^{(2)} + O(h^{3})\,,
\label{eq: alphas small h expansion}\\
\beta   &=  \beta^{(1)} + \beta^{(2)} + O(h^{3})\,,
\label{eq:betas small h expansion}
\end{align}
\end{subequations}
where the superscripts indicate the power of~$h$,
$\alpha^{(0)} = \diag(G_1,G_2,\cdots)$
and each $G_j$ has unit magnitude.
For cavity worldtubes grafted from inertial and uniformly accelerated
  segments, this situation arises when the acceleration at the centre of
  the cavity is proportional to $h/\delta$ by a numerical coefficient
  that may differ from segment to segment: the parameter $h$ is in this
  case the product of the cavity's width $\delta$ and the acceleration
  at the centre of the cavity \cite{bruschifuenteslouko11}.
We then work perturbatively in~$h$. It follows that to order
$h^2$ we have
$N=1-\tfrac{1}{4}\sum_{p,q}\bigl|V_{pq}^{(1)}\bigr|^{2}$
and
\begin{align}
\left|\,0\,\right\rangle    &=  \Bigl(1-\tfrac{1}{4}\sum\limits_{pq}\bigl|V_{pq}^{(1)}\bigr|^{2}\Bigr)\left|\,\tilde{0}\,\right\rangle
    +   \tfrac{1}{2}\sum\limits_{pq}V_{pq}\tilde{a}_{p}^{\dagger}\tilde{a}_{q}^{\dagger}\left|\,\tilde{0}\,\right\rangle
    \nonumber\\
    &\ +\tfrac{1}{8}\sum\limits_{p q i j}V_{pq}^{(1)}V_{ij}^{(1)}\tilde{a}_{p}^{\dagger}\tilde{a}_{q}^{\dagger}
        \tilde{a}_{i}^{\dagger}\tilde{a}_{j}^{\dagger}\left|\,\tilde{0}\,\right\rangle + O(h^{3})\,.
\label{eq:bosonic vacuum to second order}
\end{align}

As a first example, we take the in-region state
to be the in-vacuum~$\left|\,0\,\right\rangle$.
To order~$h^2$, the partially transposed, reduced density
matrix vanishes outside a $6\times6$ block.
Among the six eigenvalues,
the only possibly negative ones are
\begin{subequations}
\label{eq:bosonic vac par trans eigenvalues}
\begin{align}
&\lambda_{4}
= - \bigl|\beta_{kk^{\prime}}^{(1)}\bigr|^{2},
\label{eq:bosonic vac par trans eigenvalues 2,3,4}
\\
&\lambda_{6} = f^{\beta}_{k\lnot k^{\prime}} + f^{\beta}_{k^{\prime}\lnot k}
 - \left( \bigl(  f^{\beta}_{k\lnot k^{\prime}} - f^{\beta}_{k^{\prime}\lnot k} \bigr)^{2}
                    +|V_{kk^{\prime}}|^{2}\right)^{1/2},
\label{eq:bosonic vac par trans eigenvalues 5,6}
\end{align}
\end{subequations}
where $f^{\beta}_{m\lnot n}:=\tfrac{1}{2}\sum_{q\neq n}\bigl|\beta_{qm}^{(1)}\bigr|^{2}$
and $V_{kk^{\prime}}$ is kept to order~$h^2$.
$\lambda_4$ arises from coherence between $\left|\,\tilde{0}\,\right\rangle$
and $\left|\,\tilde{1}_{k}\,\right\rangle\left|\,\tilde{1}_{k^{\prime}}\right\rangle$,
while $\lambda_6$ arises from coherence between $\left|\,\tilde{0}\,\right\rangle$
and $\left|\,\tilde{2}_{k}\,\right\rangle\left|\,\tilde{2}_{k^{\prime}}\right\rangle$.

\begin{figure}
\centering
\vspace*{-4ex}%
(a)
\includegraphics[width=0.45\textwidth]{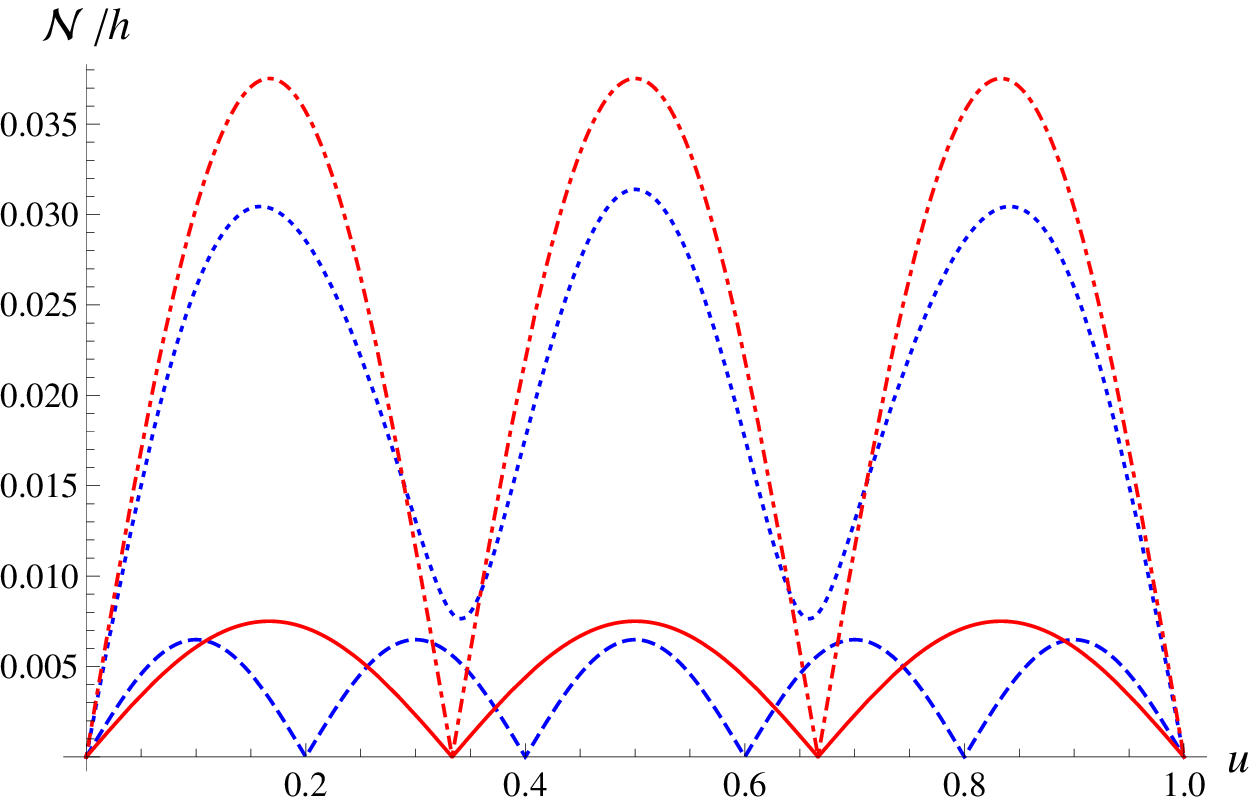}
(b)
\includegraphics[width=0.45\textwidth]{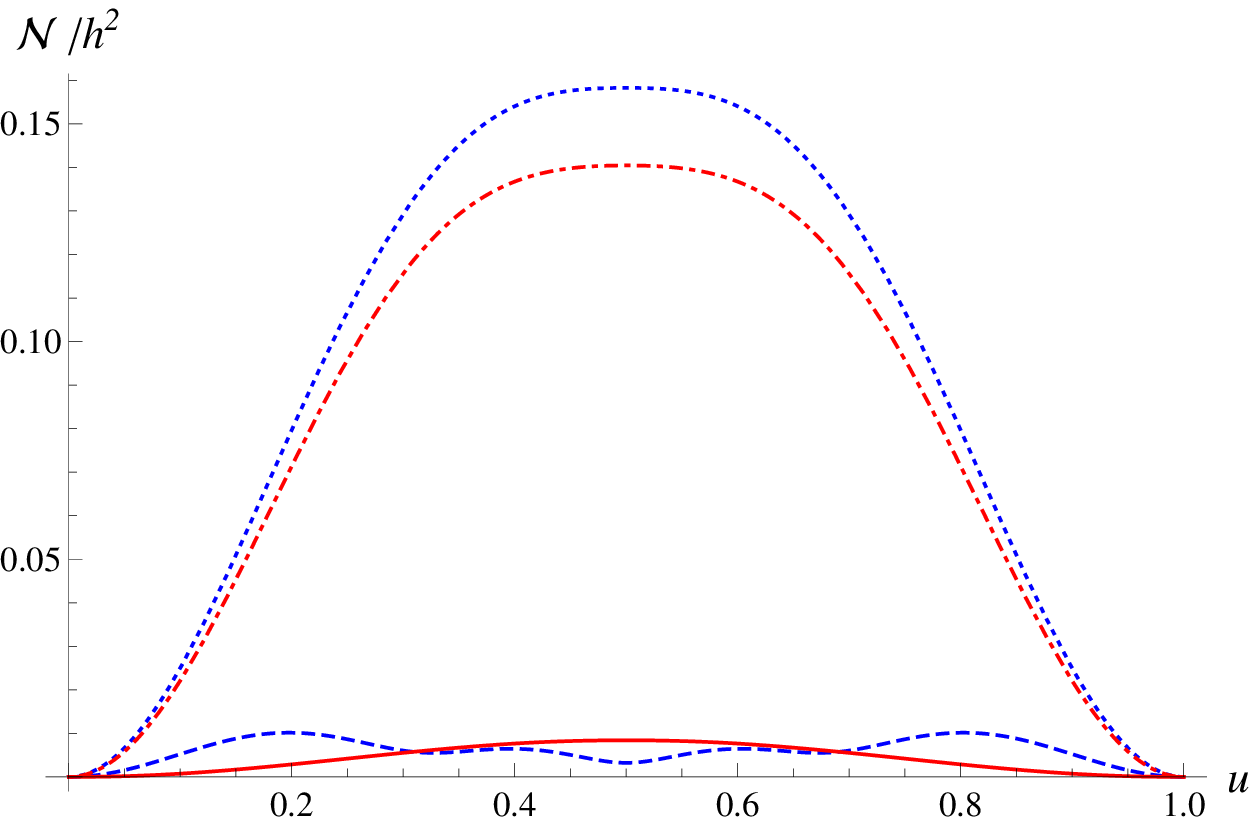}
\caption{
The leading order contribution to the negativity is shown for a
massless scalar field and a massless Dirac field. The travel scenario
has a single accelerated segment, of acceleration
$h/\delta$ as measured at the cavity's centre and of duration
$\tau$ in the cavity's proper time, where the corresponding Bogogliubov coefficients can be found in Refs.~\cite{bruschifuenteslouko11,friisleebruschilouko11}.
The negativity is periodic in $u:=h\tau/[4\delta\atanh(h/2)]$ with period~$1$.
Fig.~\ref{fig:BasicBuildingBlock}(a) shows $\mathcal{N}/h$,
in dashed (blue) for a scalar field with in-region vacuum
and $(k,k') = (1,4)$,
in dotted (blue) for a scalar field with in-region state
$\left|\,1_k\,\right\rangle$ and
$(k,k') = (1,4)$, in solid (red) for a Dirac field with in-region vacuum and
$(\kappa,\kappa') = (2,-1)$ with $s=0$,
and in dotted-dashed (red) for a Dirac field with in-region state
$\left|\!\left|\, 1_\kappa\,\right\rangle\!\right\rangle$ and
$(\kappa,\hat{\kappa}) = (1,4)$ with $s=0$
in the notation of~\cite{friisleebruschilouko11}.
Fig.~\ref{fig:BasicBuildingBlock}(b) shows the corresponding curves
for $\mathcal{N}/h^2$ with the scalar field modes $(k,k') = (1,3)$
and the fermionic modes $(\kappa,\kappa') = (1,-1)$ and
$(\kappa,\hat{\kappa}) = (1,3)$.
\label{fig:BasicBuildingBlock}}
\end{figure}

Specialising to a cavity worldtube that is grafted from inertial and
uniformly-accelerated segments, we find that a qualitative difference
emerges depending on the relative parity of $k$ and~$k'$.  If $k$ and
$k^{\prime}$ have opposite parity, the expansions given in
Ref.~\cite{bruschifuenteslouko11} and their massive generalisations show
that $\beta_{kk^{\prime}}^{(1)}$ is nonvanishing but
$V_{kk^{\prime}}^{(2)}=0$.  It follows that
$|V_{kk^{\prime}}|^{2}=\bigl|\beta_{kk^{\prime}}^{(1)}\bigr|^{2}+O(h^{4})$.
The leading term in the negativity is then \emph{linear\/} in $h$
and given by~$|\beta_{kk^{\prime}}^{(1)}|$.  If, by contrast, $k$ and
$k^{\prime}$ have the same parity, we have
$\beta_{kk^{\prime}}^{(1)}=0$ and $V_{kk^{\prime}} =
V_{kk^{\prime}}^{(2)} +O(h^3)$.  The leading term in the negativity
comes then from $\lambda_6$ and is of order~$h^2$. Sample negativity
plots for both cases are
shown in Fig.\ \ref{fig:BasicBuildingBlock} for a massless field
when the cavity undergoes a single segment of uniform acceleration.


As a second example, we take the in-region state to
be~$\left|\,1_{k}\,\right\rangle$, containing exactly one in-particle.
Using (\ref{eq:bosonic vacuum to second order}) and the
adjoint of~(\ref{eq:Bogo transformation operators out to in region}), we find
\begin{align}
\left|\,1_{k}\,\right\rangle    &=  \sum\limits_{m} \Bigl(\alpha_{mk}^{*} + \sum\limits_{p}\beta_{pk}^{(1)}V_{pm}^{(1)}
\nonumber\\
& \hspace{8ex}
- \tfrac14 \delta_{mk} G_{k}^{*} \sum\limits_{pq}\bigl|V_{pq}^{(1)}\bigr|^{2}\Bigr)\tilde{a}_{m}^{\dagger}\left|\,\tilde{0}\,\right\rangle
    \nonumber\\
    &\ +   \tfrac{1}{2}\sum\limits_{mpq} \bigl(\alpha_{mk}^{*} + G_{k}^{*}\delta_{mk}\bigr)V_{pq}
    \tilde{a}_{m}^{\dagger}\tilde{a}_{p}^{\dagger}\tilde{a}_{q}^{\dagger}\left|\,\tilde{0}\,\right\rangle
    \nonumber\\
    &\ +\tfrac18 G_{k}^{*}\sum\limits_{pqij}V_{pq}V_{ij}\tilde{a}_{k}^{\dagger}\tilde{a}_{p}^{\dagger}\tilde{a}_{q}^{\dagger}
        \tilde{a}_{i}^{\dagger}\tilde{a}_{j}^{\dagger}\left|\,\tilde{0}\,\right\rangle + O(h^{3})\,.
\label{eq:bosonic single particle to second order}
\end{align}
To order~$h^2$, the partially transposed, reduced density
matrix now vanishes outside an $8\times8$ block. Among the first five
eigenvalues, the only possibly negative one is
\begin{align}
\mu_3 = - \sqrt{3}\,\bigl|\beta_{kk^{\prime}}^{(1)}\bigr|^{2}\,,
\label{eq:bosonic single particle par trans eigenvalues 2,3}
\end{align}
which arises from coherence between
$\left|\,\tilde{1}_{k}\,\right\rangle$ and
$\left|\,\tilde{3}_{k}\,\right\rangle\left|\,\tilde{2}_{k^{\prime}}\,\right\rangle$.
The last three eigenvalues are the roots of a cubic polynomial,
analytically cumbersome for generic values of the parameters
but readily amenable to numerical work.

Specialising to a cavity worldtube that is grafted from inertial and
uniformly-accelerated segments, we again find a qualitative difference
depending on the relative parity of $k$ and~$k'$.
In particular, if $k$ and
$k^{\prime}$ have opposite parity, the leading contribution to
negativity comes from the eigenvalue
\begin{align}
\mu_{8} =  - \sqrt{\bigl|\alpha_{kk^{\prime}}^{(1)}\bigr|^{2}
+  2\bigl|\beta_{kk^{\prime}}^{(1)}\bigr|^{2}}
\label{eq:bosonic single particle par trans eigenvalues 7,8}
\end{align}
and is \emph{linear\/} in~$h$.
The negativity is in this case higher than the corresponding negativity
for the in-region state
$\left|\,0\,\right\rangle$.
Sample negativity plots are
shown in Fig.\ \ref{fig:BasicBuildingBlock} for a massless field
when the cavity undergoes a single segment of uniform acceleration.

\textbf{Fermions}:
We consider in the cavity a Dirac field $\psi$ with mass $m\ge0$, with
boundary conditions that ensure unitarity of the time
evolution as in Ref.~\cite{friisleebruschilouko11}.

Let $\left\{\psi_{n} \mid n \in\BbbZ\right\}$ be a complete
orthonormal set of mode solutions, such that the modes $n\ge0$
are of positive frequency and the modes with $n<0$ are of negative frequency
with respect to the cavity's proper time in the in-region.
We write the in-region mode expansion of the field as
$\psi =  \sum_{n\geq0}\psi_{n}b_{n}\,+\,\sum_{n<0}\psi_{n}c_{n}^{\dagger}$,
so that the nonvanishing anticommutators of the
fermionic ladder operators for positive frequency modes and negative frequency modes are
$\left\{b_{m},b_{n}^{\dagger}\right\}=\delta_{mn}$
and
$\left\{c_{m},c_{n}^{\dagger}\right\}=\delta_{mn}$ respectively and the in-vacuum
$\left|\!\left|\,0\,\right\rangle\!\right\rangle$ satisfies
$b_n \left|\!\left|\,0\,\right\rangle\!\right\rangle = 0$ and
$c_n \left|\!\left|\,0\,\right\rangle\!\right\rangle =0$.
The corresponding modes, operators and states in the out-region are denoted by a tilde.

We write the Bogoliubov transformation between the two sets of modes as
$\tilde{\psi}_{m} = \sum_{m}A_{mn}\,\psi_{n}$, where the Bogoliubov coefficients $A_{mn}$ form a unitary matrix. The relations between the two sets of ladder operators can be written in terms of the Bogoliubov
coefficients by taking appropriate inner products.
The vacua are related by
$\left|\!\left|\,0\,\right\rangle\!\right\rangle=M\,e^{\mathcal{W}}|\!|\,\tilde{0}\,\rangle\!\rangle$, where
$\mathcal{W}=\sum_{\genfrac{}{}{0pt}{3}{p\ge0}{q<0}}\,\mathcal{V}_{pq}\,\tilde{b}_{p}^{\dagger}\,\tilde{c}_{q}^{\dagger}$, $M$ is a normalisation constant, and
$\mathcal{V}_{pq}$ can be expressed in terms of $A_{mn}$~\cite{friisleebruschilouko11}.

We assume again that the Bogoliubov coefficients have a
Maclaurin expansion, now of the form
\begin{align}
A_{mn}  =  A_{mn}^{(0)} + A_{mn}^{(1)} + A_{mn}^{(2)} + O(h^{3})\,,
\end{align}
where $h$ is the small dimensionless parameter,
$A_{mn}^{(0)} = G_m \delta_{mn}$ (no sum) and
each $G_j$ has unit magnitude.
We then work perturbatively in $h$~\cite{friisleebruschilouko11}.


We first take the in-region state to be the in-vacuum $\left|\!\left|\,0\,\right\rangle\!\right\rangle$ and trace over all out-region modes except a mode $\kappa\ge0$, of positive charge,
and a mode $\kappa^{\prime}<0$, of negative charge.
As pointed out
in Ref.~\cite{monteromartinmartinez11},
the tracing in a fermionic Fock space has an ambiguity,
but the construction of
$\mathcal{V}$ guarantees that this ambiguity does not affect our measures of entanglement, given that these
measures are unambiguous for the in-region state~$\left|\!\left|\,0\,\right\rangle\!\right\rangle$. To order~$h^2$, we find that the only potentially negative eigenvalue of the partially transposed, reduced density matrix is
\begin{align}
\nu_{3}=\bar{f}_{\kappa\lnot\kappa^{\prime}}^{A} + f_{\kappa^{\prime}\lnot\kappa}^{A}
    - \Bigl(\bigl(\bar{f}_{\kappa\lnot\kappa^{\prime}}^{A} - f_{\kappa^{\prime}\lnot\kappa}^{A}\bigr)^{2}  +
    |\mathcal{V}_{\kappa\kappa^{\prime}}|^{2} \Bigr)^{1/2}
\label{eq:fermionic vac part transp eigenvalues 2,3}
\end{align}
where $\bar{f}^{A}_{m\lnot n} :=\tfrac{1}{2}\sum_{\genfrac{}{}{0pt}{3}{q<0}{q\ne n}}\bigl|A_{qm}^{(1)}\bigr|^{2}$.
Note the similarity with~(\ref{eq:bosonic vac par trans eigenvalues 5,6}).

Specialising to the massless fermion and a cavity worldtube that is grafted from inertial and uniformly-accelerated segments~\cite{friisleebruschilouko11}, we find that the eigenvalue (\ref{eq:fermionic vac part transp eigenvalues 2,3}) is linear in $h$ when $\kappa$ and $\kappa^{\prime}$ have
opposite parity. The leading order correction to the negativity is then equal to $|A^{(1)}_{\kappa\kappa^{\prime}}|$.

Consider then in-region states with particles. When the in-region state contains a single particle in mode $\kappa\geq0$, the out-region reduced density matrix for two modes of opposite charge turns out to have vanishing negativity to order~$h^2$. However, when considering two modes of the same charge as the excitation, e.g., $\kappa,\hat{\kappa}\geq0$, entanglement is generated between these modes. Mirroring the vacuum case an analogous expression to Eq.~(\ref{eq:fermionic vac part transp eigenvalues 2,3}), with $\bar{f}_{\kappa\lnot\kappa^{\prime}}^{A}$ and $f_{\kappa^{\prime}\lnot\kappa}^{A}$ replaced by $f_{\kappa\lnot\hat{\kappa}}^{A}$ and $\bar{f}_{\hat{\kappa}}^{A}$, is obtained for the possibly negative eigenvalue of the partially transposed, reduced density matrix. Similar as before the leading order correction to the negativity is $|A^{(1)}_{\kappa\hat{\kappa}}|$ if $\kappa$ and $\hat{\kappa}$ have opposite parity.

This structure is simply a consequence of the fermionic algebra and charge conservation: entanglement is generated by the Bogoliubov coefficients connecting two modes, either by creation of a pair of oppositely charged particles in these modes, or by shifting a preexistent excitation to another mode of the same charge. If the mode in question is populated, the Pauli exclusion principle prohibits any further excitation of this mode and the creation of particle--antiparticle pairs cannot entangle it with any mode of opposite charge.

When the in-region state contains a pair of oppositely charged particles, the situation changes again: the partial transpose of the out-region reduced density matrix of the corresponding modes $\kappa\geq0$ and $\kappa^{\prime}\leq0$ then has one potentially negative eigenvalue of the form of Eq.~(\ref{eq:fermionic vac part transp eigenvalues 2,3}), where $\bar{f}_{\kappa\lnot\kappa^{\prime}}^{A}$ and $f_{\kappa^{\prime}\lnot\kappa}^{A}$ are replaced by $f_{\kappa}^{A}$ and $\bar{f}_{\kappa^{\prime}}^{A}$.
Specialising to the massless fermion and a cavity worldtube that is grafted from inertial and uniformly-accelerated segments~\cite{friisleebruschilouko11}, we find that when
$\kappa$ and $\kappa^{\prime}$ have
opposite parity, the leading order correction to the negativity is the same as when the in-region state is the in-region vacuum.

Sample plots are shown in Fig.~\ref{fig:BasicBuildingBlock}.

\textbf{Conclusions}:\ We have demonstrated that nonuniform motion of
a cavity generates entanglement between modes of a quantum field confined
to the cavity, both bosonic and fermionic. Working to quadratic order in
the cavity's acceleration, and quantifying the entanglement by the
negativity, we found that the entanglement generation depends on the
initial state of the field, on the relative parity of the mode pair that
is observed at late times, and on the bosonic versus fermionic statistics.
For both bosons and fermions, we found situations where the
entanglement generation can be enhanced by placing particles in the initial
state. For fermions, however, charge conservation and the
Pauli exclusion principle require the choice of the considered
out-region modes to be consistent with the initial state to generate
entanglement, while the bosonic statistics allow the modes to be freely
populated without hindering entanglement generation.

Compared with the motion-induced entanglement degradation between a
static cavity and a moving
cavity~\cite{bruschifuenteslouko11,friisleebruschilouko11}, we found
that the entanglement generation can occur already in linear order in
the cavity's acceleration, while the entanglement degradation is a
second-order effect.  The prospects of experimental verification
\cite{bruschifuenteslouko11} could hence be significantly better for
phenomena signalling entanglement generation than entanglement
degradation.

The motion-induced entanglement effects that we have analysed have
technical similarities with the creation of squeezed states in
resonators with oscillating walls, known as the dynamical Casimir
effect~\cite{dodonovCasimirReview2010,dodonovklimovmanko90}.  In this
context, we emphasise that our only approximation was to work in the
small acceleration regime, meaning that the product of the cavity's
length and acceleration is small compared with the speed of light
squared~\cite{bruschifuenteslouko11,friisleebruschilouko11}. Our
analysis hence covers as a special case cavities that oscillate
rapidly with a small amplitude: such cavities are often introduced in
theoretical analyses of the dynamical Casimir effect but are
experimentally problematic~\cite{dodonovCasimirReview2010}.

Our analysis however covers also cavities that accelerate in a given
direction for finite but arbitrarily long times, with travel distances
that may be arbitrarily large.
Further, as the equivalence principle implies that
gravitational acceleration can be locally modelled
by acceleration in Minkowski space-time,
our results suggest that a gravitational field can produce entanglement.
Experiments for entanglement generation could
hence be sought in setups that span macroscopic
distances, including quantum communication through near-Earth satellite orbits.

\begin{acknowledgments}
We thank Gerardo Adesso, Andrzej Dragan, Marcus Huber,
Antony~R. Lee and the $\chi$-QEN collaboration for helpful
discussions and comments.
N.~F. and I.~F. acknowledge support from EPSRC
(CAF Grant No.~EP/G00496X/2 to I.~F.).
J.~L. was supported in part by STFC.
\end{acknowledgments}

\end{document}